\documentclass[a4paper]{article}
\usepackage {amssymb,amsmath,latexsym,enumerate,graphicx,epsfig,cite}
\usepackage{indentfirst}
\usepackage{titlesec}
\usepackage {subfigure,epsfig}

\usepackage{caption2}

\makeatletter\renewcommand{\@biblabel}[1]{#1.} \makeatother
\usepackage{amsthm} 

\begin{document}
\title{Exact solutions of systems of nonlinear differential equations describing the evolution of interacting populations.}
\author{Olga Yu. Efimova}
\date{} \maketitle
\begin{abstract}
The generalization of the simplest equation method to look for exact solutions of systems of nonlinear differential equations is presented. The exact solutions of NDE systems describing the evolution of two interacting populations in two cases (both populations have the low critical density or low critical density is typical for only one of populations) are obtained.
\end{abstract}

\section*{Introduction.}

Systems of nonlinear differential equations describe a lot of processes in different fields of science. Since XIX century it became clear that nonlinear phenomena are no less important than the linear ones. Therefore, the theory of nonlinear differential equations began its development.

Usually we use the approximate methods to solve the problems with the systems of nonlinear differential equations, but exact solutions are very useful too.
Exact analytical solutions allow to determine the features of solutions behavior under some initial and boundary conditions and verify the numerical calculations.

In \cite{Kudryashov2005} the simplest equation method to look for exact solutions of nonlinear differential equations was presented (see also \cite{Kudryashov2007,Kudryashov2008,Kudryashov2008a,Kudryashov2008b}). It summarizes a number of methods that were developed earlier, including the tanh-function method \cite{Lou1991,Kudryashov1996,Fan2000,Elwakil2002}, the Jacobi elliptic function
method \cite{Liu2001,Yan2003} and the Weierstrass function method \cite{Kudryashov1990, Porubov1993}.

Here we generalize the simplest equation method for systems of ordinary differential equations. This method applies two main concepts: 1) the idea of the simplest nonlinear differential equation that has lesser order than the equations of the system studied and 2)  accounting of possible singularities of the solutions of system studied.

The outline of this paper is as follows. The generalization of the simplest equation method to look for exact solutions of systems of nonlinear differential equations is given in section 1. Sections 2 and 3 are devoted to the exact solutions of NDE systems describing the evolution of two interacting populations in two cases 1) if both populations have the low critical density (section 2) and 2) if low critical density is typical for only one of populations (section 3) \cite{Bazykin2003en}.

\section{The simplest equation method to look for exact solutions of systems of nonlinear differential equations.}

Consider the system of $m$ nonlinear differential equations in the polynomial form
\begin{equation}\label{e:trs_sys}
M_i[y_1,y_2,\ldots y_m]=0,\qquad i=1,2,\ldots n
\end{equation}
and the simplest equation
\begin{equation}\label{e:trs_simp_eq}
E[Y]=0
\end{equation}
with order lesser than the ones of equations $M_i$. The solution of simplest equation is assumed to be known.

For example, we can use as the simplest equation the Riccati equation,
the equation for the Jacobi elliptic function,
the equation for the Weierstrass elliptic function,
and so on.


The unknown functions $y_j$, $j=1,2,\ldots m$ are supposed to be expressed via the solution of the simplest equation
 $Y$.

Let us describe the simplest equation method to look for exact solutions of systems of nonlinear differential equations. It contains three basic steps.

At the first step we determine the singularity orders of unknown functions $y_j$, $j=1,2,\ldots m$ by the analysis of the dominant terms of system \eqref{e:trs_sys} (this step coincides with the  first step of the Painleve test).

At the second step the solutions of ODE system \eqref{e:trs_sys} are represented as selected simplest equation solution expansions, usually in the polynomial form
 \begin{equation}\label{e:trs_podst_ob_v}
y_j= F_j(Y) = \sum_k {A_{jk} f_{k}(Y)}
\end{equation}
 The terms used in expansions are selected in consideration of singularity orders of simplest equation \eqref{e:trs_simp_eq} and singularity orders of the studied system solutions, that were determined at the first step.

 At the third step we substitute expansion \eqref{e:trs_podst_ob_v} into the system studied. After consideration of used simplest equation properties, coefficients of  various Y-function powers are set equal to zero. So we obtain the algebraic system, which solution determine coefficients $A_{jk}$ and, if it is necessary, the restrictions to the parameters of initial system.

Further we consider the applications of the simplest equation method to the systems of nonlinear differential equations that are used in ecology at the description of interacting species.

\section{Exact solutions of the nonlinear differential equations system describing the evolution of two interacting species with low critical densities.}

Evolution of two interaction populations with low critical densities can be described by the reaction-diffusion system in the form
\begin{equation}\label{e:trs_eq1}
\begin{aligned}
\frac{d u}{d t} = D_1\frac{d^2 u}{d x^2}+\alpha_1 u(K_1-u)(u-L_1)+E_1 -\varepsilon_1 uv\\
\frac{d v}{d t} = D_2\frac{d^2 v}{d x^2}+\alpha_2 v(K_2-v)(v-L_2)+E_2 -\varepsilon_2 uv
\end{aligned}
\end{equation}
Here $u\equiv u(x,t)$ and $v\equiv v(x,t)$ are the densities of populations, $t$ means time, $x$ is spatial variable.

 Type of interactions between species is determined by signs of parameters $\varepsilon_i$ (competition at $\varepsilon_1,\varepsilon_2>0$, symbiosis at $\varepsilon_1,\varepsilon_2<0$, ''predator-prey'' or ''parasite-host'' at $\varepsilon_1 \varepsilon_2<0$).

Parameters $D_i$, $\alpha_i$,  $K_i$, $L_i$ and $E_i$ are determined for each population separately and describe the evolution of population without species interactions ($i=1,\,2$). Coefficient $D_i$ characterizes the velocity of  population diffusion by influence of random motion \cite{Skellam1991}, $\alpha_i$ define the growth rate,  $K_i$ is the environment capacity, $L_i$ is the low critical density (if population is under minimum viable level it become extinct), $E_i$ describe the external influence on ecosystem \cite{Bazykin2003en, Riznichenko2002en, Riznichenko2004en}.

System of equations \eqref{e:trs_eq1} can be reduced to the Burgers-Huxley equation \cite{Estevez1990,Efimova2004en,Kudryashov2004en} by linear substitution if and only if the diffusion velocities of both populations coincide, i.e. $D_1=D_2$,   and two additional conditions on parameters of system \eqref{e:trs_eq1} hold. However, it is important to have exact solutions for different diffusion coefficients.

Substituting the variables
\begin{equation*}u'=\sqrt{\frac{2{C_0}^2}{\alpha_1 D_1 }}u,
\qquad v'=\sqrt{\frac{2{C_0}^2D_2}{\alpha_2 D_1^2}}v, \qquad z=\frac{C_0}{D_1}(x-C_0t) \end{equation*}
and denoting
\begin{gather*}
p_1=\sqrt{\frac{2\alpha_1D_1}{{C_0}^2}}(L_1+K_1),\qquad p_2=\frac{\alpha_1D_1K_1L_1}{{C_0}^2}, \qquad {E_1}'=E_1\sqrt{\frac{\alpha_1{D_1}^3}{2{C_0}^6}},\\
q_1=\sqrt{\frac{2\alpha_2 D_1^2}{{C_0}^2 D_2}}(L_2+K_2),
\qquad q_2=\frac{\alpha_2D_1^2K_2L_2}{{C_0}^2 D_2},
\qquad {E_2}'=E_2\sqrt{\frac{\alpha_2{D_1}^5}{2{C_0}^6{D_2}^3}},\\
d = \frac{D_1}{D_2}, \qquad \varepsilon_1'=\varepsilon_1\sqrt{\frac{2{D_2}}{\alpha_2{C_0}^2}}, \qquad
\varepsilon_2'=\varepsilon_2\sqrt{\frac{2{D_1}^3}{\alpha_1{C_0}^2{D_2^2}}}
\end{gather*}
omitting strokes we obtain the dimensionless form of system \eqref{e:trs_eq1} in traveling wave variables
\begin{equation}\label{e:trs_eq1-ob}
\begin{gathered}
\frac{d^2 u}{d z^2} +\frac{d u}{d z} - 2 u^3 + p_1 u^2 -p_2 u +E_1 -\varepsilon_1 uv =0 \hfil \\
\frac{d^2 v}{d z^2} + d \frac{d v}{d z} - 2 v^3 + q_1 v^2 -q_2 v+E_2 -\varepsilon_2 uv = 0 \hfil
\end{gathered}
\end{equation}

Use the simplest equation method.

The dominant terms of system \eqref{e:trs_eq1-ob} are the following
 \begin{equation}
\begin{aligned}
\frac{d^2 u}{d z^2} -2 u^3=0\\
\frac{d^2 u}{d z^2} - 2 v^3=0
\end{aligned}
\end{equation}
This truncated system has solution $u(z)=s_1/z$, $v(z)=s_2/z$, where $s_1$ and $s_2$ can take on values $\pm1$ independently. Therefore, solutions of system \eqref{e:trs_eq1-ob} have first order singularities.

As the simplest equation we use the Riccati equation in the form
\begin{equation}\label{e:trs_Riccati}
\frac{d Y}{d z}=-Y^2+P_1Y+P_0
\end{equation}
where $Y\equiv Y(z)$ and $P_0$, $P_1$ are constants.

The Riccati equation belongs to the class of exactly solvable ones, its solutions have first order singularities.
Nonhomogeneous solutions of equation \eqref{e:trs_Riccati} can be written as
\begin{itemize}
\item At $P_1^2+4P_0>0$
\begin{gather}
Y(z)=\frac{1}{2}\left( P_1+\sqrt{P_1^2+4P_0}\tanh\bigl(\sqrt{P_1^2+4P_0}(z-z_0)\bigr) \right) \label{e:trs_solRicc1}
\\ Y(z)=\frac{1}{2}\left( P_1+\sqrt{P_1^2+4P_0}\coth\bigl(\sqrt{P_1^2+4P_0}(z-z_0)\bigr) \right)
\end{gather}

\item At $P_1^2+4P_0=0$
\begin{equation}
Y(z)=\frac{P_1}{2} +\frac{1}{z-z_0}
\end{equation}

\item At $P_1^2+4P_0<0$
\begin{gather}
Y(z)=\frac{1}{2}\left( P_1-\sqrt{-(P_1^2+4P_0)}\tan\bigl(\sqrt{-(P_1^2+4P_0)}(z-z_0)\bigr) \right)
\\ Y(z)=\frac{1}{2}\left( P_1+\sqrt{-(P_1^2+4P_0)}\cot\bigl(\sqrt{-(P_1^2+4P_0)}(z-z_0)\bigr) \right)
\end{gather}
\end{itemize}

Here $z_0$ is the constant of integration. Note, that  function \eqref{e:trs_solRicc1} is the only one bounded  on complete number scale, however other solutions can be used on bounded intervals.

As solutions of system \eqref{e:trs_eq1-ob} are to have first order singularities, we use the substitution
\begin{equation*}
\begin{aligned}
u(z)=\widetilde{A}_0Y+\widetilde{A}_1+\widetilde{A}_2\frac{Y_z}{Y}\\
v(z)=\widetilde{B}_0Y+\widetilde{B}_1+\widetilde{B}_2 \frac{Y_z}{Y}
\end{aligned}
\end{equation*}
Taking into the account the properties of function $Y$ ( i.e. the Riccati equation \eqref{e:trs_Riccati}), the equivalent form of this substitution is
\begin{equation}\label{e:trs_podst1}
\begin{aligned}
u(z)=A_0Y+A_1+\frac{A_2 P_0}{Y}\\
v(z)=B_0Y+B_1+\frac{B_2 P_0}{Y}
\end{aligned}
\end{equation}

Nonhomogeneity condition for solution \eqref{e:trs_podst1} is
\begin{equation}\label{e:trs_usl_neodn1}
A_0^2+A_2^2P_0^2 \ne 0, \qquad    B_0^2+B_2^2P_0^2 \ne 0
\end{equation}

\emph{Statement.}

We can assume $A_0\ne0$ without loss of generality.

\emph{Proof.}
Suppose that $A_0=0$. If $A_2P_0=0$ then the nonhomogeneity condition \eqref{e:trs_usl_neodn1} fails, therefore  $P_0\ne0$, $A_2\ne0$. Denote $\widehat{Y}=P_0/Y$.
\begin{multline}\widehat{Y}_z=(P_0/Y)_z =-\frac{P_0}{Y^2}Y_z=-\frac{P_0}{Y^2}(-Y^2+P_1Y+P_0)
=\\=-\frac{P_0^2}{Y^2}-P_1\frac{P_0}{Y}+P_0=-\widehat{Y}^2+\widetilde{P}_1 \widehat{Y} + P_0\end{multline}
so function $\widehat{Y}$ satisfy the Riccati equation in the form \eqref{e:trs_Riccati} with $\widetilde{P}_1=-P_1$.
Therefore solution \eqref{e:trs_podst1} at $A_0=0$ can be presented as
\begin{equation*}
\begin{aligned}
&u(z)=\widehat{A}_0 \widehat{Y}+A_1\\
&v(z)=\widehat{B}_0\widehat{Y}+B_1+\frac{\widehat{B}_2}{\widehat{Y}}
\end{aligned}
\end{equation*}
where $\widehat{A}_0=A_2$, $\widehat{B}_0=B_2$ and $\widehat{B}_2=B_0$. \qed

Substituting \eqref{e:trs_podst1} in system \eqref{e:trs_eq1-ob}, taking into acctount
\begin{equation}
\begin{aligned}
Y_z =-Y^2+P_1Y+P_0\\
Y_{zz}=2Y^3-3P_1y^2+(P_1^2-2P_0)Y+P_0P_1
\end{aligned}
\end{equation}
and setting coefficients of  various $Y$-function powers equal to zero, we get the system of fourteen algebraic equations. Consider its solutions.

Equations generated by the largest and the least powers of $Y$ are
\begin{equation*}
A_0(A_0^2-1)=0,\quad A_2(A_2^2-1)=0, \quad B_0(B_0^2-1)=0, \quad B_2(B_2^2-1)=0
\end{equation*}
so we can sort possible solutions by values of parameters $A_0$, $A_2$, $B_0$ and $B_2$.
\begin{itemize}
\item Case $A_2=B_2=0$, $A_0^2=B_0^2=1$.

Here substitution \eqref{e:trs_podst1} is invariant under linear transformations, so we can consider $P_1=0$ without loss of generality.
Therefore 
we define the parameters of solution
\begin{equation}
\begin{gathered}
A_1= \frac{ A_{{0}}p_{{1}}-B_{{0}}\varepsilon_{{1}}-1
 }{6A_{0}} \\
B_1= \frac{B_{{0}}q_{{1}} -A_{{0}}\varepsilon_{{2}}-d
 }{6B_{{0}}} \\
P_0=-1/12\,\varepsilon_{{1}} \left( -\varepsilon_{{2}}+p_{{1}} \right) B_{{0}}A_
{{0}}+1/12\,\varepsilon_{{1}} \left( d-1 \right) B_{{0}}+{}\hfil\\ \hfil{}+1/12\,{p_{{1}}}^
{2}-1/12\,\varepsilon_{{1}}q_{{1}}-1/12-1/2\,p_{{2}}
\end{gathered}
\end{equation}
and coefficients of initial system, that allow this solution
\begin{equation}
\begin{gathered}
q_2=1/6\,{q_{{1}}}^{2}-1/6\,{p_{{1}}}^{2}+1/6\,\varepsilon_{{1}}q_{{1}}+p_{{2
}}+1/6-1/6\,\varepsilon_{{2}}p_{{1}}-1/6\,{d}^{2} -{}\hfil\\\hfil{}-1/6\,\varepsilon_{{2}}
 \left( d-1 \right) A_{{0}}-1/6\,\varepsilon_{{1}} \left( d-1 \right) B_{
{0}}+ \left( 1/6\,\varepsilon_{{1}}p_{{1}}-1/6\,q_{{1}}\varepsilon_{{2}}
 \right) B_{{0}}A_{{0}}\\
 E_1= A_{{1}} \left( 2\,{A_{{1}}}
^{2}-p_{{1}}A_{{1}}+B_{{1}}\varepsilon_{{1}}+p_{{2}} \right)-  A_{{0}}P_{{0}}\\
E_2=B_{{1}} \left( 2\,{B_{{1}}}
^{2}-q_{{1}}B_{{1}}+\varepsilon_{{2}}A_{{1}} +q_{{2}}\right)-d B_{{0}}P_{{0}}
\end{gathered}
\end{equation}

\item Case $A_2=B_0=0$, $A_0^2=B_2^2=1$.

Here system \eqref{e:trs_eq1-ob} has solution in the form \eqref{e:trs_podst1} with
\begin{equation}
\begin{gathered}
A_1=B_1=0\\
P_1=(p_1A_0-1)/3\\
P_0=(p_{{1}}A_{{0}}-2+{p_{{1}}}^{2}-9\,p_{{2}})/18
\end{gathered}
\end{equation}
at condition that
\begin{equation}
\begin{gathered}
q_1=B_2(1+d-p_1A_0)
\\
q_2=(1+d+3\,p_{{2}}-p_{{1}} \left( d+1 \right) A_{{0}})/3
\\
E_1=P_{{0}}  A_{{0}} \left( \varepsilon_{{1}}B_{{2}}-P_{{1}}-1 \right)
\\
E_2=P_{{0}}B_{{2}} \left( \varepsilon_{{2}}A_{{0}} + P_{{1}}-d\right)
\end{gathered}
\end{equation}

\item Case $A_2=0$,  $A_0^2=B_0^2=B_2^2=1$.

Here
\begin{equation}
\begin{gathered}
A_1=0\\
P_1=(p_{{1}}A_{{0}}-1-\varepsilon_{{1}}B_{{0}})/3 \\
B_1=(q_{{1}}+B_{{2}}(3P_{{1}}-d))/6\\
P_0=(P_{{1}}+{P_{{1}}}^{2}-p_{{2}}-\varepsilon_{{1}}B_{{1}})/2\\
\varepsilon_2 = A_{{0}} ( d(B_0B_2-1)-3P_1(1+B_0B_2))
\\
E_1=A_{{0}}P_{{0}}(\varepsilon_{{1}}B_{{2}}-P_1-1)
\\
E_2=P_{{0}} \left(  \left( -2\,q_{{1}}+\varepsilon_{{2}}A_{{0}}B_{{0}}+12\,B_{{1}} \right) B_{{0}}B_{{2}}-d \left( B_{{0}}+B_{{2}}
 \right) -P_{{1}} \left( B_{{0}}-B_{{2}} \right)  \right) +{}\hfill \\\hfil{}+2\,{B_{{1}}
}^{3}+q_{{2}}B_{{1}}-q_{{1}}{B_{{1}}}^{2}
\\
q_2=P_1^2-6B_1^2+2B_1q_1-dP_1-2P_0(1+3B_0B_2)
\\
q_1=B_{{2}}(d-3P_{{1}})+12\,{{P_{{1}}d A_{{0}}B_{{0}}}/\varepsilon_{{2}}}
\end{gathered}
\end{equation}

\item Case  $B_2=0$,  $A_0^2=A_2^2=B_0^2=1$.

Here
\begin{equation}
\begin{gathered}
B_1=0\\
P_1=(q_{{1}}B_{{0}}-d-\varepsilon_{{2}}A_{{0}})/3 \\
A_1=(p_{{1}}+A_{{2}}(3P_{{1}}-1))/6\\
P_0=(dP_{{1}}+{P_{{1}}}^{2}-q_{{2}}-\varepsilon_{{2}}A_{{1}})/2\\
\varepsilon_1 = B_{{0}} (A_0A_2-1-3P_1(1+A_0A_2))
\\
E_1=P_{{0}} \left(  \left( -2\,p_{{1}}+\varepsilon_{{2}}A_{{0}}B_{{0}}+12\,A_{{1}} \right) A_{{0}}A_{{2}}- \left( A_{{0}}+A_{{2}}
 \right) -P_{{1}} \left( A_{{0}}-A_{{2}} \right)  \right) +{}\hfill \\\hfil{}+2\,{A_{{1}}
}^{3}+p_{{2}}A_{{1}}-p_{{1}}{A_{{1}}}^{2}
\\
E_2=B_{{0}}P_{{0}}(\varepsilon_{{2}}A_{{2}}-P_1-d)
\\
p_2=P_1^2-6A_1^2+2A_1p_1-P_1-2P_0(1+3A_0A_2)
\\
p_1=A_{{2}}(1-3P_{{1}})+12\,{{P_{{1}}d A_{{0}}B_{{0}}}/\varepsilon_{{1}}}
\end{gathered}
\end{equation}

\end{itemize}

Other solutions of algebraic system either don't satisfy the conditions of positiveness of $p_i$, $q_i$ and $d$ or reduce to solutions sited above.

\begin{figure}[t]
\center%
\epsfig{file=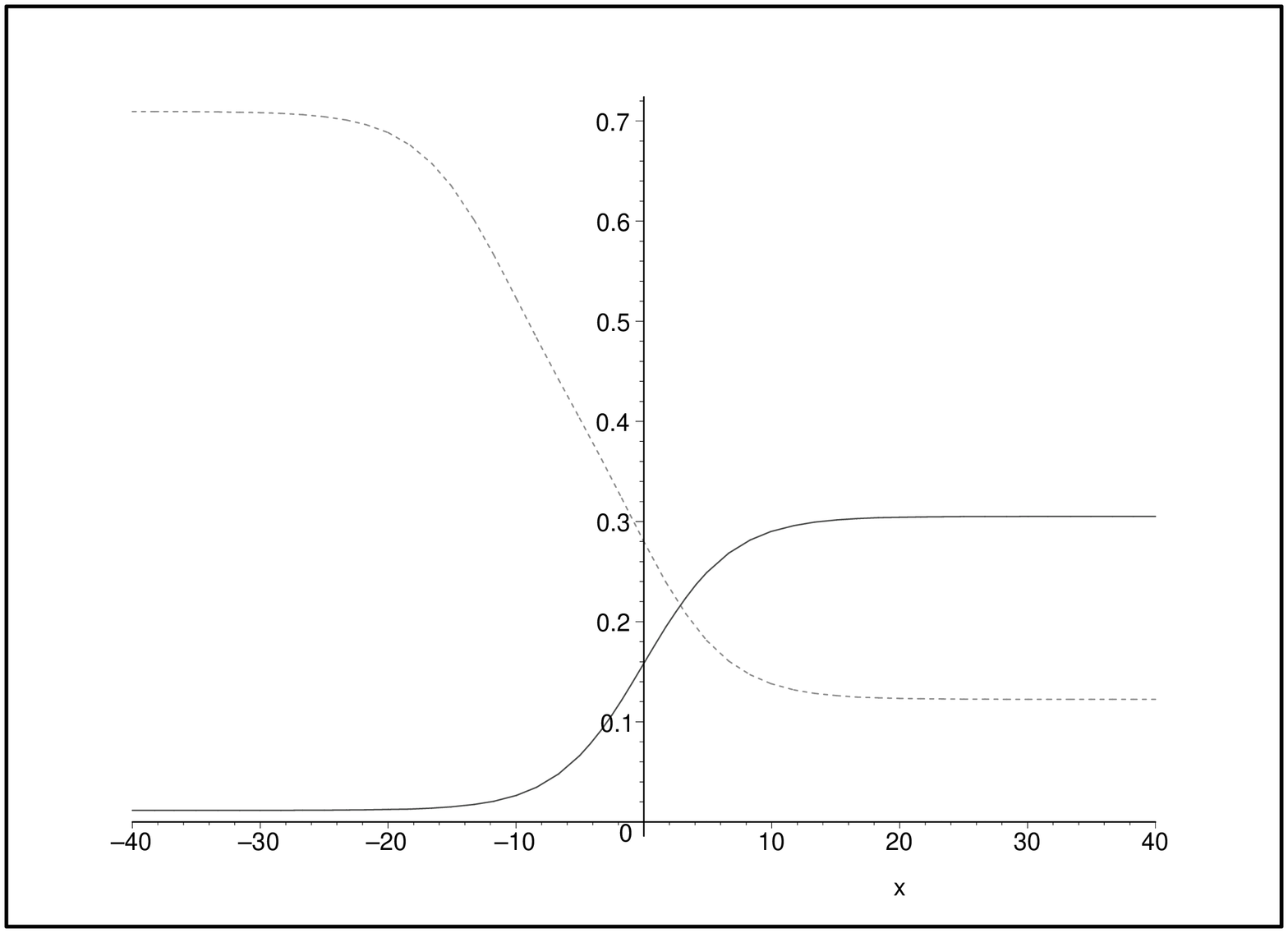,width=70mm,angle=-90}
\caption{}
\label{fig:trs-f2}
\end{figure}

\begin{figure}[t]
\center%
\epsfig{file=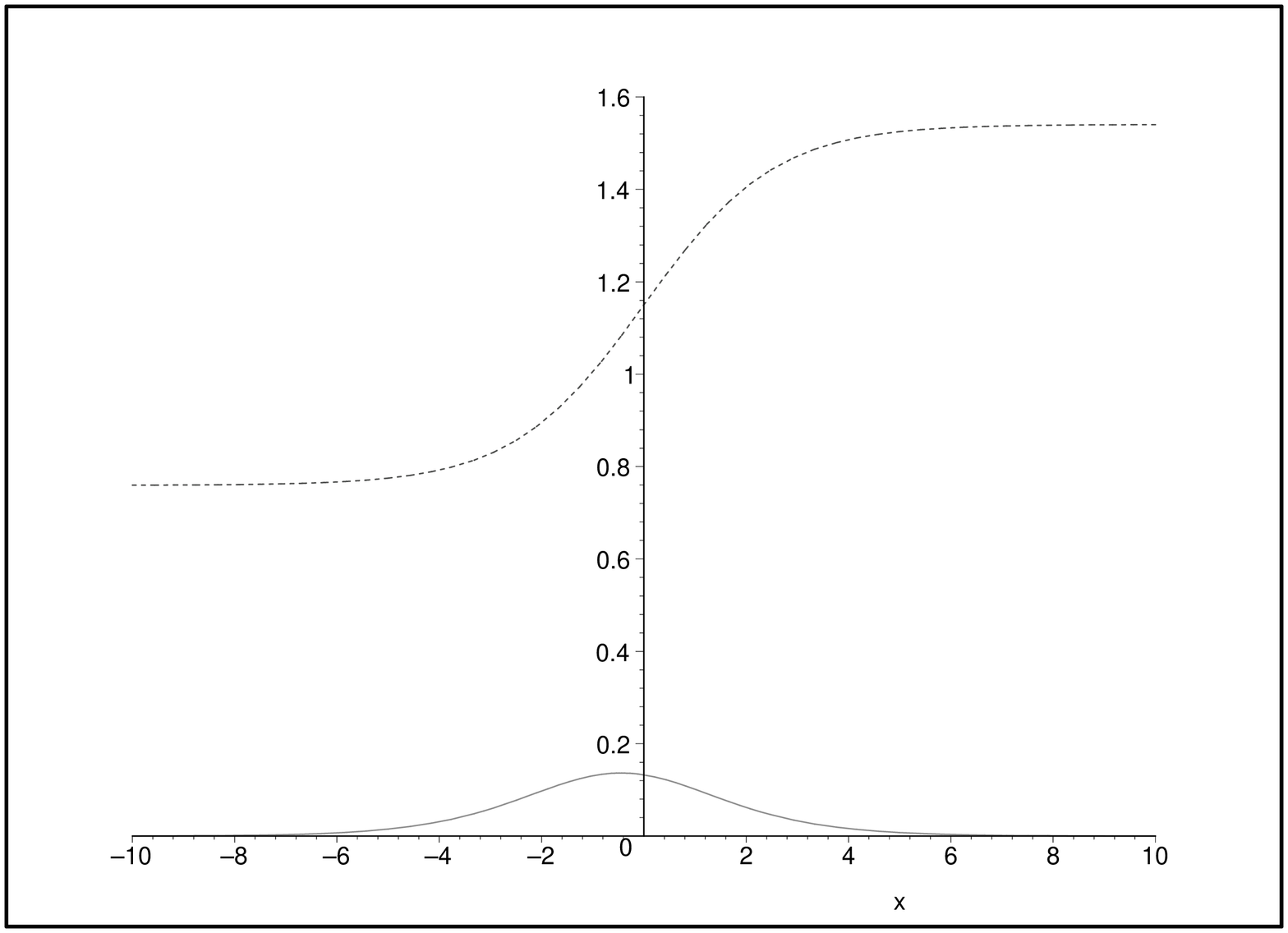,width=70mm,angle=-90}
\caption{}
\label{fig:trs-f1}
\end{figure}

If we consider solutions bounded  on complete number scale only, the solutions obtained describe two types of populations behavior: 1) simultaneous monotone change of both populations densities  (fig. \ref{fig:trs-f2})
and 2) change of one of populations state by act of the interactions with the solitary wave of the other population (fig. \ref{fig:trs-f1})

Note, that behavior of the second type is generated by solutions in case 3 at  $B_2=-B_0$ and in case 4 at $A_2=-A_0$.
Thus solution given on fig. \ref{fig:trs-f1} is obtained at $p_1=4.8$, $p_2=2.8$, $d=0.8$, $\varepsilon_1=3.1$, $q_1=7.7$, $q_2=2.45$, $\varepsilon_2=-1.6$, $E_1=0.234$, $E_2=0$ and has the form

\begin{align*}
&u(z)=\frac{23}{20}+\frac{\sqrt{61}}{20}\tanh\left({\frac{\sqrt{61}}{20}(z-z_0)}\right),\\ &v(z)=\frac{23}{20}-\frac{\sqrt{61}}{20}\tanh\left({\frac{\sqrt{61}}{20}(z-z_0)}\right)-
\frac{117}{5\left(23+\sqrt{61}\tanh\left({\frac{\sqrt{61}}{20}(z-z_0)}\right)\right)} \end{align*}
where $z_0$ is the arbitrary constant (at fig. \ref{fig:trs-f1} it assumed to be zero).
It corresponds to the situation when the predators rise results in the prey extinction, but predators can subsist in the bounded domain only.

\section{Exact solutions of the nonlinear differential equations system describing the evolution of two interacting species, one of which is characterized by low critical density.}

Consider the system of nonlinear differential equations  in the form
\begin{equation}\label{e:trs_eq2}
\begin{gathered}
\frac{d u}{d t} = D_1\frac{d^2 u}{d x^2}+2 \alpha_1 u(K_1-u)(u-L_1)+E_1 -\varepsilon_1 uv\\
\frac{d v}{d t} = D_2\frac{d^2 v}{d x^2}+\alpha_2 v(K_2-v)+E_2 -\varepsilon_2 uv
\end{gathered}
\end{equation}
In contrast to system \eqref{e:trs_eq1}, it describes the situation, when only one of populations is characterized by low critical density, and the other one  in the absence of interacting species and external influence complies with the Ferhulst-Pearl equation \cite{Pearl1930, Riznichenko2002en,Bazykin2003en, Riznichenko2004en}. The meanings of parameters of system coincide with the ones declared in section 2.

Substituting the variables
\begin{equation*}
u'=2(K_1+L_1)u,
\qquad
v'=2\frac{\alpha_1 D_2}{\alpha_2 D_1}(K_1+L_1)^2v,
\qquad z={\sqrt{\frac{D_1}{2\alpha_1}}} \frac{(x-C_0t)}{(K_1+L_1)}
\end{equation*}
and denoting
\begin{gather*}
p= \frac{K_1L_1}{2(K_1+L_1)^2},\qquad
{E_1}'=\frac{E_1}{4\alpha_1(L_1+K_1)^3},\qquad
{\varepsilon_1}'=\varepsilon_1 \frac{D_2}{D_1\alpha_2},\\
q=\frac{\alpha_2D_1}{\alpha_1D_2(L_1+K_1)^2}K_2,\qquad
{E_2}'=\frac{{D_1}^2\alpha_2E_2}{4\alpha_1^2D_2^2(K_1+L_1)^4},\qquad
d=\frac{D_1}{D_2},\\
C_1=\frac{C_0}{\sqrt{2\alpha_1D_1}(K_1+L_1)},\qquad
{\varepsilon_2}'=\varepsilon_2 \frac{D_1}{\alpha_1D_2(K_1+L_1)}
\end{gather*}
omitting strokes we obtain the dimensionless form of system \eqref{e:trs_eq2} in traveling waves variables
\begin{equation}\label{e:trs_eq2-ob}
\begin{aligned}
\frac{d^2 u}{d z^2}+C_1 \frac{d u}{d z} -2u^3+u^2-p u+E_1 -\varepsilon_1 uv=0\\
\frac{d^2 v}{d z^2} + d C_1 \frac{d v}{d z}-v^2 + q v +E_2 -\varepsilon_2 uv=0
\end{aligned}
\end{equation}

The dominant terms of system \eqref{e:trs_eq2-ob} are the following
 \begin{equation}
\begin{aligned}
\frac{d^2 u}{d z^2} -2 u^3 -\varepsilon_1 uv=0\\
\frac{d^2 u}{d z^2} - 2 v^2=0
\end{aligned}
\end{equation}
The solution of this truncated system is $u(z)=s_1\sqrt{1-3\varepsilon_1}/z$, $v(z)=6/z^2$, where  $s_1$ can take on value $\pm1$.

So the solution of system \eqref{e:trs_eq2-ob} we are looking in the form
\begin{equation}\label{e:trs_podst2}
\begin{aligned}
u(z)=A_0Y+A_1+\frac{A_2 P_0}{Y}\\
v(z)=B_0Y^2+B_1Y+B_2+\frac{B_3 P_0}{Y}+\frac{B_4{P_0}^2}{Y^2}
\end{aligned}
\end{equation}
where $Y\equiv Y(z)$ is the solution of the Riccati equation \eqref{e:trs_Riccati} (here we take into account that function $u$ has first order singularities and function $v$ has the second order singularities).

The nonhomogeneity condition for solution in the form \eqref{e:trs_podst2} is the following
\begin{equation}\label{e:trs_usl_neodn2}
{A_0}^2+{A_2}^2{P_0}^2 \ne 0, \qquad    {B_0}^2+{B_4}^2{P_0}^4 \ne 0
\end{equation}

Substituting transformation \eqref{e:trs_podst2} in system \eqref{e:trs_eq2-ob}, taking into account the Riccati equation \eqref{e:trs_Riccati} and setting coefficients of various $Y$-function powers equal to zero, we obtain sixteen algebraic equations. Equations generated by the largest and the least powers of $Y$ are
\begin{equation*}
\begin{aligned}
A_0(2-2{A_0}^2-\varepsilon_1 B_0)=0\\
A_2{P_0}^3 (2-2{A_2}^2-\varepsilon_1 B_4)=0\\
B_0(B_0-6)=0\\
B_4 {P_0}^4 (B_4-6)=0
\end{aligned}
\end{equation*}

We can sort three cases by values of parameters $A_0$, $A_2$, $B_0$ and $B_4$, which specify all possible solutions of system \eqref{e:trs_eq2-ob} in the form \eqref{e:trs_podst2} under natural conditions of parameters of initial system.
Consider them sequentially.

\begin{itemize}
\item Case  ${A_0}^2=1-3\varepsilon_1$, $B_0=6$, $A_2=B_4=0$.

As a result we have that $B_3=0$. At $A_2=B_4=0$ transformation \eqref{e:trs_podst2} is invariant under linear transformations, so we assume $P_1=0$ without loss of generality.
Then we can find sequentially determine the solution parameters
\begin{equation}
\begin{gathered}
B_1=-3(\varepsilon_2A_0+2C_1d)/5\\
A_1=\frac{ \left( 1-3\,
\varepsilon_{{1}} \right)  \left( 5+3\,\varepsilon_{{1}}\varepsilon_{{2}}
 \right)-A_{{0}} C_{{1}} \left( 5-6\,\varepsilon_{{1}}d \right) }{30(1-2\varepsilon_1)}\\
 P_0=2 \varepsilon_{{2}} A_{{0}}\frac{{\varepsilon_{{2}}}^{2}-3\,\varepsilon_{{1}}{
\varepsilon_{{2}}}^{2}+125\,\varepsilon_{{2}}A_{{1}}-125\,q+17\,{C_{{1}}}^{2
}{d}^{2} }{200(12 C_1 d+\varepsilon_2 A_0)}+{}\hfil \\ \hfill {}+3\,dC_{{1}} \frac{ 8\,{C_{{1}}}^{2}{d}^{2}+27
\,\varepsilon_{{1}}{\varepsilon_{{2}}}^{2}-9\,{\varepsilon_{{2}}}^{2}}{200(12 C_1 d+\varepsilon_2 A_0)}\\
B_2=(6\,q-{B_{{1}}}^{2}-C_{{1}}dB_{{1}}-48\,P_{{0}}
-6\,\varepsilon_{{2}}A_{{1}}-\varepsilon_{{2}}A_{{0}}B_{{1}})/12
\end{gathered}
\end{equation}
and coefficients of the initial system, at which this solution exists
\begin{equation}
\begin{gathered}
p=2\,A_{{1}}-6\,{A_{{1}}}^{2}-2\,P_{{0}}-\varepsilon_{{1}}B_{{2}}-{\frac {\varepsilon_{{1}}A_{{1}}B_{{1}}}{A_{{0}}}}\\
E_1=pA_{{1}}-{A_{{1}}}^{2}-C_{{1}}A_{{0}}P_{{0}}+\varepsilon_{{1}}A_{{1}}B_{{
2}}+2\,{A_{{1}}}^{3}\\
E_2=\varepsilon_{{2}}A_{{1}
}B_{{2}}-12\,{P_{{0}}}^{2}-C_{{1}}dB_{{1}}P_{{0}}+{B_{{2}}}^{2}-qB_{{2}}
\end{gathered}
\end{equation}

\item Case ${A_0}^2=1-3\varepsilon_1$, $A_2^2=1$, $B_0=6$, $B_4=0$.

Here the solution parameters and coefficients of system \eqref{e:trs_eq2-ob} are to satisfy the relations
\begin{equation}
\begin{gathered}
E_1=A_{{1}} \left( 2\,{A_{{1}}}^{2}-A_{{1}}+p \right) -P_{{0}} \left( P_{{
1}}+C_{{1}} \right) A_{{0}}-{}\hfil \\ \hfil{}-P_{{0}} \left( -\epsilon_{{1}}B_{{1}}+C_{{
1}}-P_{{1}} \right) A_{{2}}+2\,P_{{0}} \left( 6\,A_{{1}}-1 \right) A_{
{2}}A_{{0}}
\\
E_2=\epsilon_{{2}}A_{{2}}P_{{0}}B_{{1}}-12\,{P_{{0}}}^{2}-B_{{1}}P_{{1}}P_
{{0}}-C_{{1}}dB_{{1}}P_{{0}}
\\
p=-6\,A_{{0}}A_{{2}}P_{{0}}-2\,P_{{0}}-C_{{1}}P_{{1}}-6\,{A_{{1}}}^{2}+{
P_{{1}}}^{2}+2\,A_{{1}}
\\
q=-4\,{P_{{1}}}^{2}+1/2\,B_{{1}}P_{{1}}+\varepsilon_{{2}}A_{{1}}+1/6\,{B_{{
1}}}^{2}+1/6\,C_{{1}}dB_{{1}}+{}\hfil\\ \hfil{}+1/6\,\varepsilon_{{2}}A_{{0}}B_{{1}}-2\,C_{
{1}}dP_{{1}}+8\,P_{{0}}
\\
P_0=-\frac {B_{{1}} \left( -6\,C_{{1}}dP_{{1}}-18\,{P_{{1}}}^{2}+3
\,B_{{1}}P_{{1}}+{B_{{1}}}^{2}+C_{{1}}dB_{{1}}+\varepsilon_{{2}}A_{{0}}B_
{{1}} \right) }{36(B_{{1}}+6\,P_{{1}}-\varepsilon_{{2}}A_{{2}}+2\,C_{{1}}d)}
\\
B_1=-3/5\,\varepsilon_{{2}}A_{{0}}-6/5\,C_{{1}}d-6\,P_{{1}}
\\
A_1=1/6+ \left( 1/2\,P_{{1}}-1/6\,C_{{1}} \right) A_{{2}}
\\
B_2=B_3=0\\
P_1={\frac {C_{{1}} \left( 6\,\varepsilon_{{1}}d-5 \right) A_{{0}}-5\,
C_{{1}} \left( -1+2\,\varepsilon_{{1}} \right) A_{{2}}-\varepsilon_{{1}}
 \left( 9\,\varepsilon_{{1}}\varepsilon_{{2}}+5-3\,\varepsilon_{{2}} \right) }{
 15\left( 1-2\,\varepsilon_{{1}} \right) (A_{{0}}+ A_{{2}})}}
\end{gathered}
\end{equation}
and condition
\begin{multline}
 \left( C_{{1}}P_{{1}}-6\,{A_
{{1}}}^{2}-p+2\,A_{{1}}+{P_{{1}}}^{2}-2\,P_{{0}}
 \right) A_{{0}}-{}\hfil\\ \hfil{}-6\, \left( 1-3\,\varepsilon_{{1}} \right) A_{{2}}P_{{0}}-\varepsilon_{{1}} \left( A_{{1}}B_{{1}}+6\,A_{{2}}P_{{0}
} \right)=0
\end{multline}
(which can be reduced to the cubic equation in $C_1$) must hold.

\begin{figure}[t]
\center%
\epsfig{file=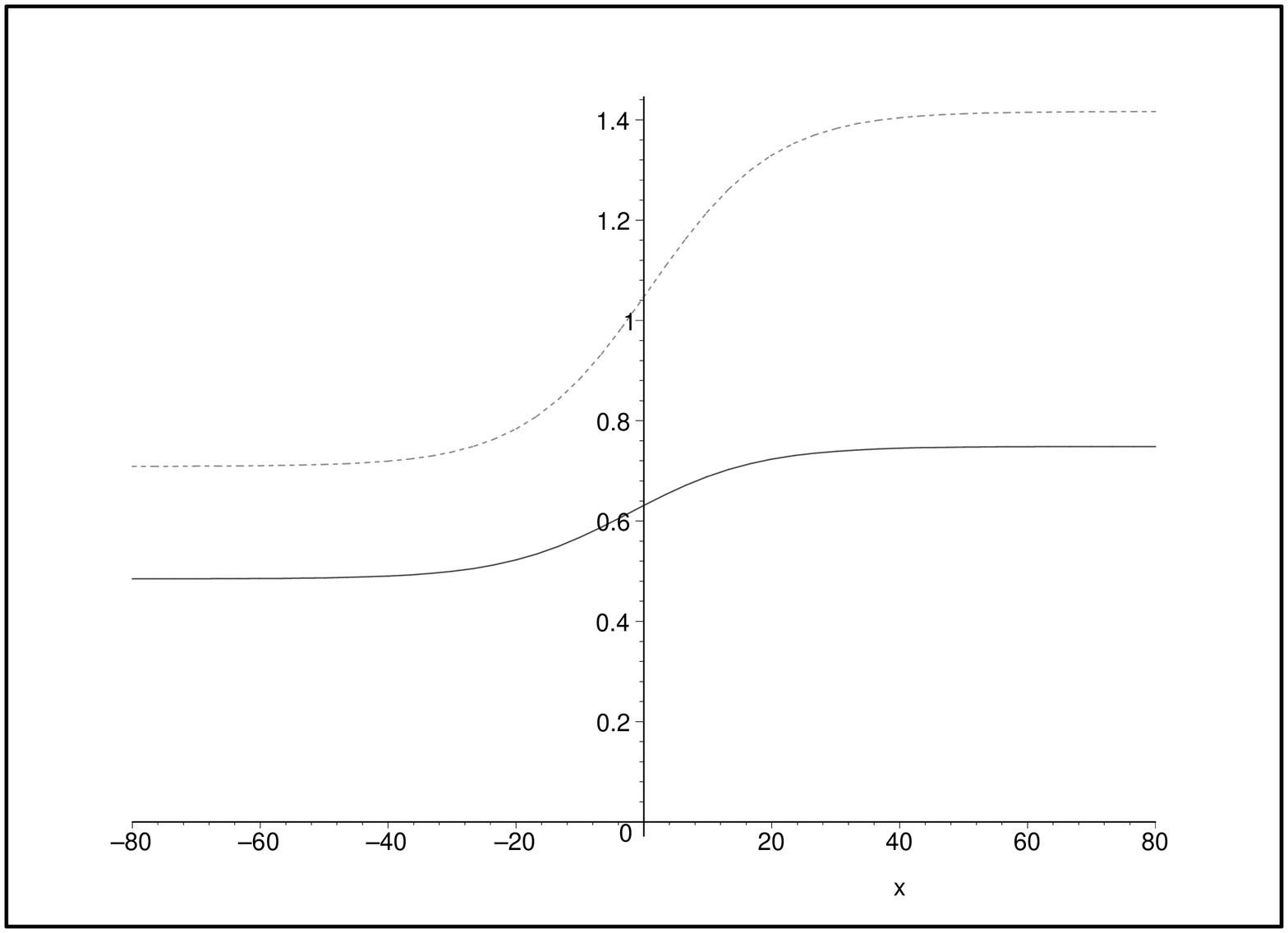,width=70mm,angle=-90}
\caption{}
\label{fig:trs-2-1}
\end{figure}

\begin{figure}[t]
\center%
\epsfig{file=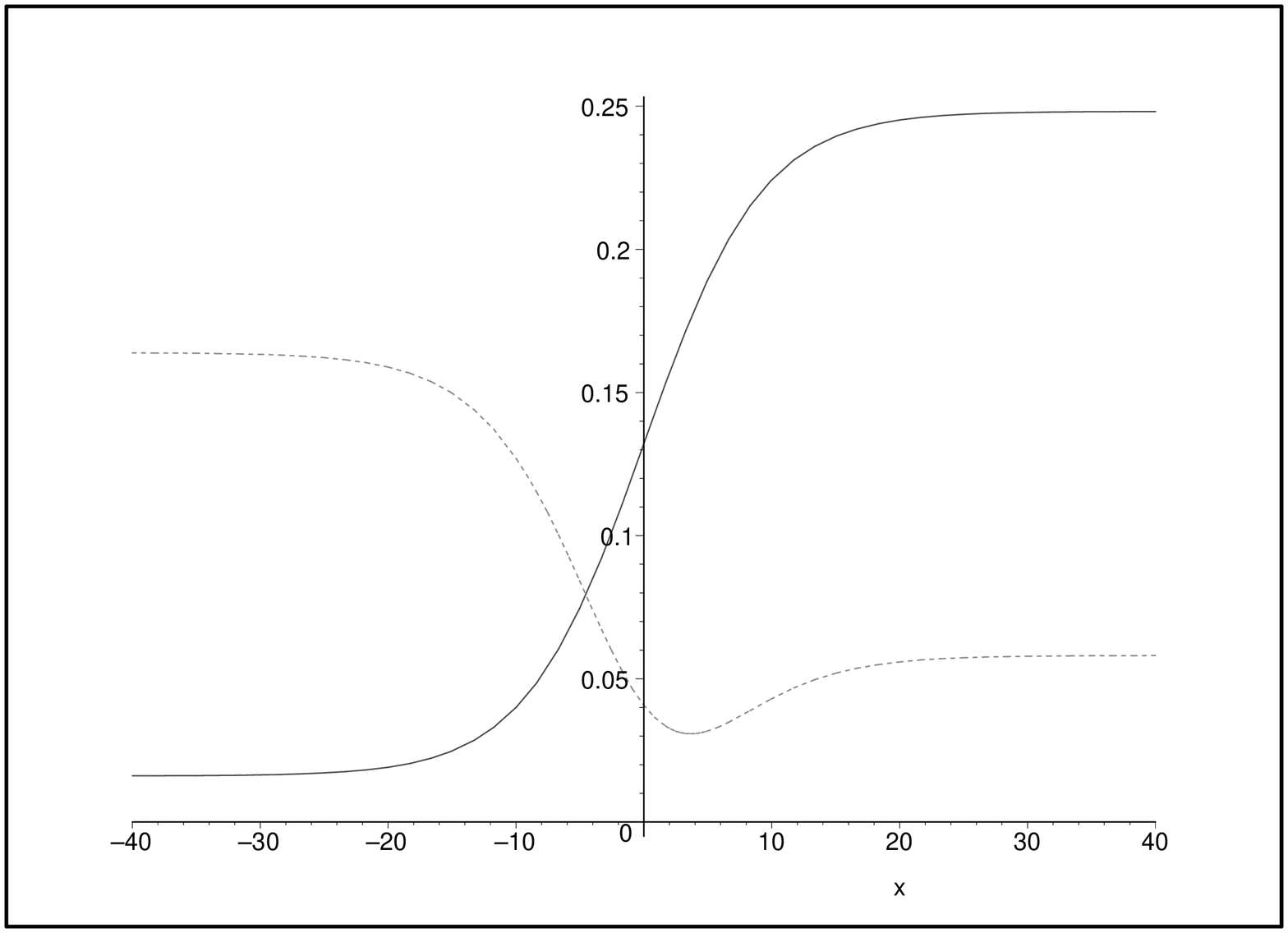,width=70mm,angle=-90}
\caption{}
\label{fig:trs-2-2}
\end{figure}

\item Case  ${A_0}^2=1-3\varepsilon_1$, $A_2=-A_0$, $B_0=B_4=6$.

Solution with parameters
\begin{equation}
\begin{gathered}
A_1=\frac{2d-5}{6(d-5)}, \qquad
B_1=-\frac{2d}{d-5}A_0-\frac{6d}{5}C_1, \\
B_2=\frac{9d^2}{200}{C_1}^2+\frac{5d(2d-5)}{16(d-5)^2},\qquad
B_3=\frac{2d}{d-5}A_0-\frac{6d}{5}C_1,\\
P_1=-\frac{d}{3(d-5)}A_0,\qquad
P_0= \frac{d^2}{400}{C_1}^2-\frac{19d(2d-5)}{288(d-5)^2}
\end{gathered}
\end{equation}
is allowed at
\begin{equation}
\begin{gathered}
p=\frac{d(4d-25)}{400}{C_1}^2-\frac{5(4d+3)(2d-5)}{96(d-5)^2},\\
q= \frac{3d^2}{20}{C_1}^2+\frac{3d(2d-5)}{8(d-5)^2},\\
\varepsilon_1=\frac{5}{6d},\qquad
\varepsilon_2=\frac{20d}{3(d-5)},\qquad E_1=0,\\
 E_2= \frac{351d^4}{40000}{C_1}^4-\frac{21d^3(2d-5)}{64(d-5)^2}{C_1}^2-\frac{9d^2(2d-5)^2}{256(d-5)^4}\\
\end{gathered}
\end{equation}
\end{itemize}

\begin{figure}[t]
\center%
\epsfig{file=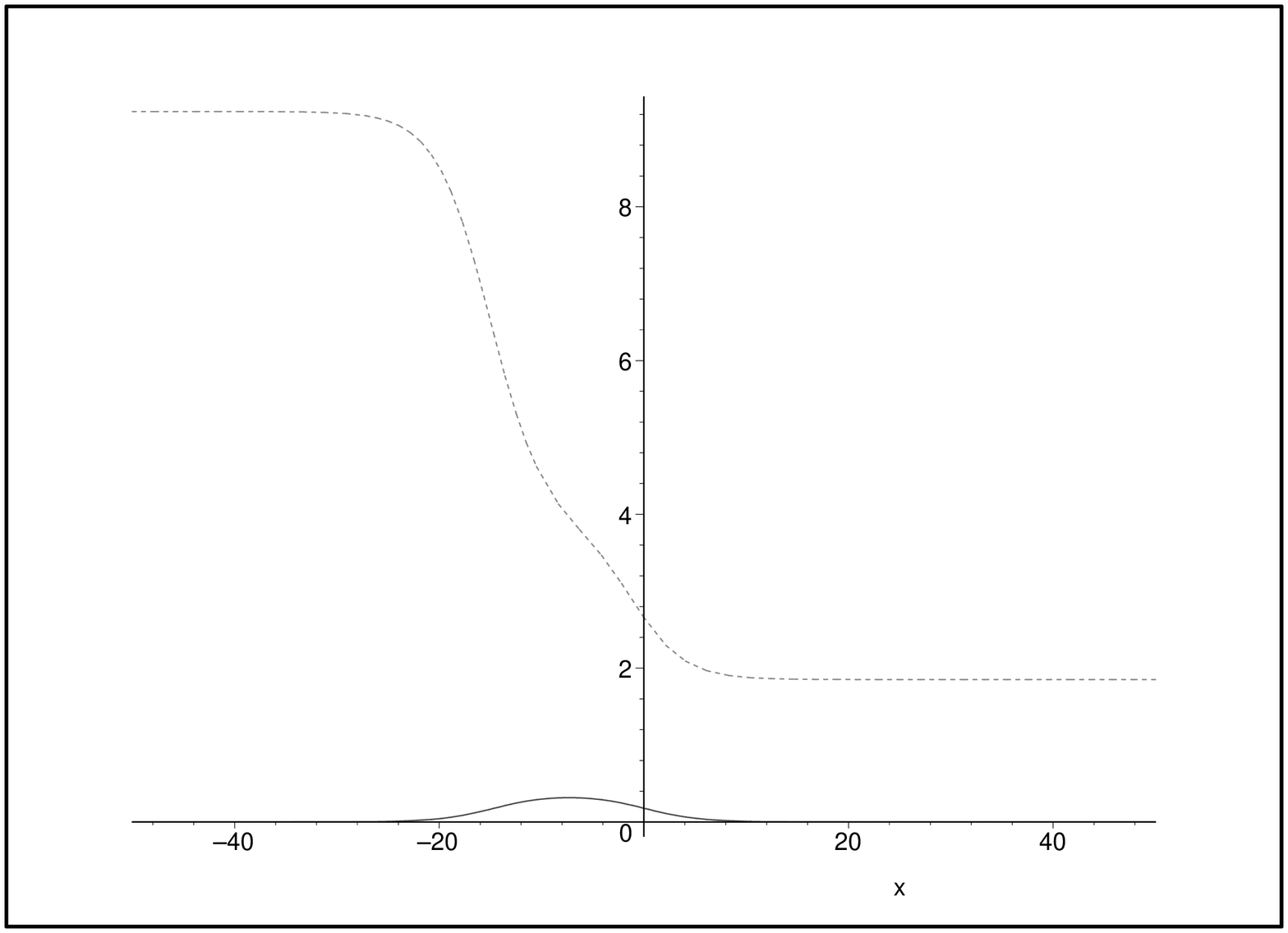,width=70mm,angle=-90}
\caption{}
\label{fig:trs-2-3}
\end{figure}

Solutions, given in this section, can demonstrate more complicated behavior in comparison with ones obtained in section 2. Some examples are presented at figures \ref{fig:trs-2-1}, \ref{fig:trs-2-2} and \ref{fig:trs-2-3}. Thus at fig. \ref{fig:trs-2-2} density of one of populations under the influence of the other population diffusion falls before the  equilibrium state mounts.

\section*{Conclusion.}

In this paper the generalization of simplest equation method to look for exact solutions of nonlinear differential equations is described.

Using the simplest equation method, we obtain the exact solutions of systems describing the evolution of two interacting species in two cases  1) when both populations are characterized by low critical densities and
  2) when low critical density is important for one of populations only and the other one follows the Ferhulst-Pearl law.

All the solutions obtained describe the spatial transitions between the steady states of the system studied.

\bibliographystyle{unsrt}     
\bibliography{my_refs}
\end{document}